# Técnicas Avanzadas de Ciberseguridad: Integración y Evolución de la Kill Chain en Diversos Escenarios

Juan Diego Bermudez, Josue Joel Castro, Diego Alejandro Peralta, Pablo Alejandro Guacaneme

*Resumen*—El documento proporciona un análisis profundo de los principales modelos de cadena de ataque utilizados en la ciberseguridad, incluyendo el marco Cyber Kill Chain de Lockheed Martin, el marco MITRE ATT&CK, el modelo Diamond y el IoTKC, enfocándose en sus fortalezas y debilidades. Posteriormente, se destaca la necesidad de una mayor adaptabilidad y exhaustividad en el análisis de ataques, lo que ha llevado a la creciente preferencia por marcos como MITRE ATT&CK y el modelo Diamond. Una revisión de los ataques internos en la computación en la nube muestra cómo la combinación de árboles de ataque y Kill Chain puede ofrecer una metodología efectiva para identificar y detectar este tipo de amenazas, centrando los esfuerzos de detección y defensa en los nodos críticos. Asimismo, se hace énfasis en la importancia de los modelos de análisis avanzados, como BACCER, en la identificación y detección de patrones de ataque y lógica de decisión utilizando técnicas de inteligencia y acciones defensivas y ofensivas.

**Palabras clave:** Cadena de Ataque, Internet de las Cosas, Seguridad en IoT, Modelo de Markov, Software SIEM, Troyanos de Acceso Remoto, Sistema SCADA

*Abstract*—The document provides an in-depth analysis of the main attack chain models used in cybersecurity, including the Lockheed Martin Cyber Kill Chain framework, the MITER ATT&CK framework, the Diamond model, and the IoTKC, focusing on their strengths and weaknesses. Subsequently, the need for greater adaptability and comprehensiveness in attack analysis is highlighted, which has led to the growing preference for frameworks such as MITRE ATT&CK and the Diamond model. A review of insider attacks in cloud computing shows how the combination of attack trees and kill chains can offer an effective methodology to identify and detect these types of threats, focusing detection and defense efforts on critical nodes. Likewise, emphasis is placed on the importance of advanced analysis models, such as BACCER, in the identification and detection of attack patterns and decision logic using intelligence techniques and defensive and offensive actions.

## I. Introducción

La ciberseguridad es un campo en constante evolución que requiere de innovaciones y enfoques cada vez más sofisticados para contrarrestar las crecientes amenazas. A medida que los actores maliciosos adoptan tácticas y técnicas más avanzadas, es imperativo que la comunidad de ciberseguridad mantenga un ritmo de innovación equivalente, o superior. En la epoca digital actual, el tema de la ciberseguridad se ha convertido en un aspecto fundamental y prioritario en todos los niveles de la sociedad, desde empresas multinacionales hasta individuos. La creciente dependencia de tecnologías y sistemas informáticos sofisticados ha traído consigo una mayor vulnerabilidad a diversas amenazas cibernéticas. En este contexto, es esencial explorar y comprender el tema de las "Técnicas Avanzadas de Ciberseguridad: Integración y Evolución de la Kill Cahin en Diversos Escenarios".

El término ciberseguridad abarca una amplia gama de prácticas, tecnologías y medidas diseñadas para proteger sistemas, redes y datos de ataques o accesos no autorizados. En la última década, los avances tecnológicos han permitido el desarrollo de técnicas de ciberseguridad más sofisticadas y robustas. Sin embargo, también han surgido nuevas y diversas formas de amenazas cibernéticas, obligando a la evolución constante del campo de la ciberseguridad.

Un componente crítico en este contexto es la Kill Chain, una serie de etapas que un ciberdelincuente puede seguir para llevar a cabo un ataque exitoso. Comprender y combatir eficazmente estas etapas es una parte esencial de cualquier estrategia de ciberseguridad. Sin embargo, con la creciente integración de tecnologías y la aparición de escenarios cada vez más diversos, la kill chain también está evolucionando.

Este trabajo pretende proporcionar una visión integral y actualizada sobre las técnicas avanzadas de ciberseguridad y cómo se están integrando y evolucionando en diferentes cadenas de ataque y escenarios. Al hacerlo, se pretende arrojar luz sobre las tácticas, técnicas y procedimientos actuales y emergentes en ciberseguridad, proporcionando una base sólida para la futura formulación de políticas y la toma de decisiones en este ámbito crítico.

## II. Antecedentes y Contexto

En el vasto y en constante cambio del paisaje digital, la importancia de la ciberseguridad nunca ha sido tan crítica. Con cada avance tecnológico que redefine nuestro modo de interacción con la información y los sistemas, se revela un nuevo espectro de amenazas y vulnerabilidades. Los ciberdelincuentes, aprovechando estas brechas, han desarrollado tácticas cada vez más sofisticadas y devastadoras. En consecuencia, la ciberseguridad ha tenido que evolucionar rápidamente, adoptando nuevos enfoques y técnicas para mantenerse al día con estas amenazas en constante cambio.

Un concepto crítico en este contexto es el de la Cadena de Ataque o "Kill Chain". Este modelo, desarrollado originalmente por Lockheed Martin, sirve como un marco estructurado para entender y describir las etapas sucesivas de un ataque cibernético. Desde el reconocimiento inicial y la infiltración hasta la explotación y exfiltración de datos, la Kill Chain permite a los profesionales de la ciberseguridad descomponer y analizar los ataques, proporcionando un enfoque sistemático para la defensa cibernética.

Sin embargo, la naturaleza fluida y dinámica de la ciberseguridad ha llevado a una evolución constante de este modelo. Varios marcos y modelos de Cadena de Ataque han surgido a lo largo de los años, cada uno diseñado para abordar diferentes tipos de amenazas y escenarios. Algunos de los más notables incluyen el marco MITRE ATT&CK, que proporciona una base de conocimientos en constante actualización de tácticas y técnicas de adversarios basadas en observaciones del mundo real, y el Diamond Model, que adopta un enfoque cuadri facético para analizar y contrarrestar los ataques cibernéticos, centrándose en el adversario, las capacidades, la infraestructura y la víctima.

Con la llegada de nuevas tecnologías y paradigmas, como el Internet de las Cosas (IoT) y la computación en la nube, la ciberseguridad ha enfrentado nuevos y significativos desafíos. Los dispositivos IoT, por ejemplo, a menudo carecen de las protecciones de seguridad adecuadas, lo que los convierte en blancos fáciles para los ciberdelincuentes. De manera similar, la computación en la nube, aunque trae consigo enormes beneficios de eficiencia y escalabilidad, también presenta riesgos de seguridad únicos debido a su naturaleza compartida y distribuida. Estos desafíos han requerido que los modelos de Cadena de Ataque se adapten y se especialicen para abordar eficazmente estas nuevas amenazas.

### III. Modelos de Cadena de Ataque

**Definición de modelo de cadena de ataque**

El modelo de cadena de ataque, conocido también como Kill Chain, es un marco conceptual crucial en el campo de la ciberseguridad que permite comprender y analizar el proceso completo que sigue un atacante durante un ciberataque. Este modelo se basa en la premisa de que los ataques cibernéticos no son eventos aislados, sino que siguen una secuencia de etapas interconectadas que los atacantes utilizan para identificar vulnerabilidades, explotarlas y lograr sus objetivos.

El modelo de cadena de ataque se divide en varias etapas clave que incluyen el reconocimiento, la entrega, la explotación, la instalación, el comando y control, y la acción. Cada etapa representa un paso en el proceso general del ataque y está diseñada para ayudar a las organizaciones a comprender las tácticas y técnicas utilizadas por los atacantes en cada fase. Al comprender y estudiar estas etapas, las organizaciones pueden fortalecer sus defensas, anticipar y prevenir ataques, y mejorar la capacidad de detección y respuesta ante amenazas cibernéticas.

El modelo de cadena de ataque es esencial para desarrollar estrategias de ciberseguridad efectivas y abordar los desafíos actuales en el panorama de amenazas. Proporciona una estructura sistemática que permite a las organizaciones comprender el proceso de ataque desde una perspectiva holística, identificar posibles puntos débiles y desarrollar contramedidas proactivas en cada etapa. Al combinar este modelo con otros enfoques y herramientas de seguridad, las organizaciones pueden mejorar su capacidad para detectar, prevenir y mitigar los ciberataques, lo que resulta en una mayor resiliencia y protección de sus activos y datos críticos.

**Modelo Kill Chain de Lockheed Martin**

El Modelo Kill Chain de Lockheed Martin es un marco conceptual ampliamente reconocido en la ciberseguridad que describe las etapas y acciones que los atacantes siguen durante un ciberataque. Se enfoca en comprender las tácticas y técnicas utilizadas en cada etapa, lo que permite anticiparse a los ataques, implementar medidas de seguridad proactivas y responder de manera más eficaz.

El Modelo Kill Chain de Lockheed Martin se ha convertido en un marco de referencia ampliamente adoptado para comprender y contrarrestar los ciberataques. Al comprender las diferentes etapas de la cadena de ataque, las organizaciones pueden implementar medidas de seguridad y defensas en cada punto de la cadena para detectar, prevenir y mitigar los ataques en curso o futuros.

El Modelo Kill Chain se divide en las siguientes etapas:

1. Reconocimiento: En esta fase, el atacante recopila información sobre el objetivo seleccionado. Esto puede incluir la identificación de sistemas, la búsqueda de información sobre empleados, la exploración de perfiles en redes sociales y la recopilación de datos relevantes para el ataque.

2. Arma: En esta etapa, el atacante adquiere o desarrolla las herramientas y exploits necesarios para llevar a cabo el ataque. Esto implica la búsqueda de vulnerabilidades conocidas o la creación de malware personalizado para aprovechar las debilidades identificadas durante la fase de reconocimiento.

3. Entrega: Aquí, el atacante proporciona el medio para entregar el ataque al objetivo. Esto puede implicar el envío de correos electrónicos de phishing, la explotación de vulnerabilidades en servicios web, la inserción de dispositivos USB maliciosos o cualquier otra técnica para engañar al objetivo y lograr que ejecute el ataque.

4. Explotación: En esta etapa, el atacante busca explotar las vulnerabilidades en el sistema objetivo. Puede incluir la ejecución de código malicioso, la explotación de vulnerabilidades de software o hardware, la manipulación de archivos o la realización de acciones que permitan al atacante ganar control sobre el sistema.

5. Instalación: Después de obtener acceso al sistema, el atacante busca establecer una presencia persistente para mantener el control incluso después de que se detecte y elimine el acceso inicial. Esto puede implicar la instalación de puertas traseras, la creación de cuentas de usuario adicionales o la modificación de configuraciones para facilitar el acceso futuro.

6. Comando y control: En esta etapa, el atacante establece una comunicación encubierta con los sistemas comprometidos. Esto permite al atacante mantener el control y enviar instrucciones adicionales. Puede implicar la configuración de servidores de comando y control (C&C) o el uso de canales encubiertos para mantener la comunicación sin ser detectado fácilmente.

7. Acción: En la última etapa del Modelo Kill Chain, el atacante lleva a cabo la acción final para lograr sus objetivos específicos. Esto puede incluir actividades como la exfiltración de datos confidenciales, el sabotaje de sistemas, el robo de información valiosa o cualquier otro objetivo que el atacante haya buscado desde el principio.

Limitaciones:

· Enfoque lineal: El modelo Kill Chain sigue un enfoque secuencial y lineal de las etapas de un ataque, desde el reconocimiento hasta la acción final. Sin embargo, los ciberataques reales pueden ser más complejos y no siempre siguen una progresión lineal. Los adversarios pueden saltar entre etapas o utilizar tácticas más sofisticadas que no se ajustan perfectamente al flujo de la cadena de ataque tradicional.

· Adaptabilidad limitada: El modelo Kill Chain se ha desarrollado como un enfoque general para describir las etapas de un ataque, pero puede no ser adecuado para todos los escenarios y tipos de ataques. Algunos ataques pueden tener variantes o características únicas que no se ajustan fácilmente a las etapas establecidas por el modelo. Esto puede requerir adaptaciones o extensiones del modelo para abordar adecuadamente esos casos específicos.

· Enfoque centrado en el adversario: El modelo Kill Chain se enfoca en comprender las tácticas y técnicas utilizadas por los adversarios. Si bien esto proporciona información valiosa para las defensas, el modelo puede no capturar adecuadamente otros aspectos importantes de la ciberseguridad, como las vulnerabilidades del sistema, los errores humanos o los controles de seguridad inadecuados. Por lo tanto, es importante complementar el modelo Kill Chain con otros enfoques y estrategias para una defensa holística.

· Evolución de las tácticas de ataque: A medida que los adversarios desarrollan nuevas tácticas y técnicas, el modelo Kill Chain puede volverse obsoleto o incompleto en términos de representar los ataques más recientes. Los ciberataques están en constante evolución, y los atacantes buscan formas de evadir las defensas tradicionales. Esto significa que el modelo Kill Chain debe actualizarse continuamente para mantenerse relevante y abordar las últimas tendencias en ciberataques.

**Modelo MITRE ATT&CK**

Otro modelo relevante en el campo de la cadena de ataque es el modelo MITRE ATT&CK (Adversarial Tactics, Techniques, and Common Knowledge). Desarrollado por MITRE Corporation, es un marco ampliamente utilizado que describe tácticas y técnicas comunes utilizadas por los atacantes en diferentes etapas de un ciberataque. Este se centra en las tácticas y técnicas utilizadas por los adversarios, proporcionando una visión detallada de sus comportamientos y capacidades. A diferencia del modelo Kill Chain, que se enfoca en el flujo secuencial de un ataque, ATT&CK adopta un enfoque más granular y desglosa las tácticas y técnicas en una matriz de referencia.

La matriz de referencia de ATT&CK se organiza en tácticas y técnicas, con múltiples categorías y subcategorías dentro de cada una. Las tácticas representan los objetivos generales de un atacante, como el reconocimiento, la explotación o la persistencia, mientras que las técnicas son acciones específicas utilizadas para lograr esas tácticas. Cada técnica en la matriz de referencia está respaldada por información detallada, que incluye descripciones, ejemplos de herramientas y mitigaciones recomendadas. Esto permite a los equipos de seguridad comprender mejor cómo se llevan a cabo los ataques en cada etapa y cómo pueden detectarlos y prevenirlos.

El modelo MITRE ATT&CK se basa en una matriz de referencia que organiza las tácticas y técnicas utilizadas por los adversarios en ciberataques. La matriz de referencia está compuesta por las siguientes categorías:

1. Tácticas: Representan los objetivos generales de un atacante durante un ciberataque. Estas tácticas se dividen en varias categorías, como Reconocimiento, Intrusión, Ejecución, Persistencia, Exfiltración, etc. Cada categoría de táctica se enfoca en una fase específica del ciclo de vida del ataque.

2. Técnicas: Son acciones específicas empleadas por los atacantes para lograr las tácticas definidas. Las técnicas se clasifican dentro de cada categoría de táctica y proporcionan una descripción detallada de cómo se llevan a cabo las acciones maliciosas. Cada técnica se identifica mediante un identificador único y se documenta con información sobre su descripción, comportamiento observado, ejemplos de herramientas y mitigaciones recomendadas.

3. Subtécnicas: Dentro de las técnicas, también se definen subtécnicas. Estas representan variantes específicas o pasos adicionales dentro de una técnica particular. Las subtécnicas permiten una mayor especificidad en la descripción de los comportamientos y tácticas utilizadas por los adversarios.

El modelo MITRE ATT&CK es ampliamente utilizado tanto en la industria como en la investigación de ciberseguridad. Ayuda a los equipos de defensa a mejorar su postura de seguridad al comprender las tácticas y técnicas utilizadas por los adversarios y proporciona un lenguaje común para comunicar y compartir información sobre amenazas.

Limitaciones:
Dentro de las limitaciones, podemos destacar las siguientes:

· Complejidad de actualización: Dado que el modelo ATT&CK se mantiene constantemente actualizado para

- reflejar las tácticas y técnicas más recientes utilizadas por los adversarios, su mantenimiento y actualización pueden ser complejos y requerir esfuerzos continuos. Esto implica que puede llevar tiempo para que las nuevas técnicas y tácticas sean documentadas y agregadas al modelo, lo que puede resultar en una brecha temporal entre las últimas amenazas emergentes y su inclusión en ATT&CK.

- Especificidad limitada: Si bien ATT&CK proporciona una amplia cobertura de tácticas y técnicas, es posible que algunas técnicas específicas o variantes particulares no estén incluidas en la matriz de referencia. Esto se debe a que el modelo no puede abarcar todas las posibles acciones maliciosas y variaciones que pueden surgir en un ciberataque. Por lo tanto, puede ser necesario recurrir a otros recursos y fuentes de información para obtener una comprensión completa de las amenazas específicas.

- Adaptabilidad a contextos específicos: ATT&CK es un modelo amplio que puede aplicarse a diversos contextos y escenarios de ciberseguridad. Sin embargo, debido a su naturaleza generalizada, es posible que sea necesario realizar adaptaciones y ajustes para que sea más relevante y aplicable a un entorno o sector específico. Esto implica la necesidad de personalizar y contextualizar el modelo según las necesidades y características particulares de cada organización o industria.

- Enfoque basado en observaciones: El modelo ATT&CK se basa en comportamientos y tácticas observadas en ataques reales y en la información disponible en la comunidad de ciberseguridad. Sin embargo, esto puede resultar en una falta de cobertura completa de todas las posibles tácticas y técnicas utilizadas por los adversarios. Nuevas técnicas innovadoras o altamente sofisticadas pueden no estar representadas en ATT&CK hasta que se detecten y documenten en la comunidad de seguridad.

3.4. Adaptación de la cadena de ataque a diferentes escenarios

El modelo de cadena de ataque, debido a su naturaleza flexible y adaptable, se puede ajustar y personalizar para adaptarse a diversos escenarios y entornos de ciberseguridad. Cada sector, industria o entorno organizacional puede presentar características únicas en términos de infraestructura, sistemas y amenazas, lo que requiere adaptaciones específicas del modelo para abordar los desafíos particulares que surgen en esas situaciones.

Una de las formas de adaptar la cadena de ataque es mediante la inclusión de etapas adicionales que sean relevantes para un escenario específico. Por ejemplo, en el ámbito de los sistemas ciberfísicos, donde se integran componentes físicos con sistemas de control, se puede agregar una etapa adicional para abordar la interacción entre estos elementos y considerar las posibles amenazas y vulnerabilidades específicas de ese entorno.

Asimismo, en sectores como el financiero, de salud o energético, donde se enfrentan amenazas específicas, es posible ajustar el modelo para abordar tácticas y técnicas que son más comunes y relevantes en esos escenarios. Esto permite una mejor comprensión de las vulnerabilidades y riesgos particulares de cada industria y facilita el desarrollo de estrategias de seguridad más efectivas y adaptadas a esas circunstancias específicas.

La adaptación de la cadena de ataque a diferentes escenarios y entornos proporciona una mayor contextualización y especificidad en la comprensión de las amenazas y los ataques. Permite a las organizaciones identificar y priorizar los riesgos específicos a los que se enfrentan, así como personalizar sus estrategias y medidas de seguridad para abordarlos de manera más efectiva. Esto incluye la selección de controles y tecnologías de seguridad específicas, la implementación de monitoreo y detección especializados, y la capacitación y concientización del personal en relación con las amenazas relevantes a su sector o entorno.

En resumen, la adaptación de la cadena de ataque a diferentes escenarios y entornos es esencial para una ciberseguridad eficaz. Al personalizar el modelo, las organizaciones pueden abordar los desafíos y riesgos específicos que enfrentan en su sector o industria, desarrollando estrategias y medidas de seguridad más efectivas y contextualizadas. Esto les permite estar mejor preparadas para enfrentar las amenazas cibernéticas y proteger sus activos críticos en su entorno específico.

IV. Cadena de Ataque en el Internet de las Cosas (IoT)

El Internet de las cosas (IoT) ha experimentado un crecimiento masivo en la última década, con dispositivos conectados que abarcan desde electrodomésticos y dispositivos de consumo hasta infraestructuras críticas y sistemas industriales. Sin embargo, con la creciente adopción de dispositivos IoT, la superficie de ataque se expande, exponiendo redes y sistemas a amenazas y brechas de seguridad cada vez más sofisticadas. En este contexto, es fundamental comprender la cadena de ataque en el IoT, una secuencia de eventos que un ciberdelincuente debe realizar para comprometer con éxito un sistema o red de IoT.

La cadena de ataque, o modelo Kill Chain, se originó en el ámbito militar y fue adaptada al ciberespacio por Lockheed Martin. Tradicionalmente, consta de siete fases: reconocimiento, armamento, entrega, explotación, instalación, comando y control, y acciones sobre los objetivos. Este modelo ha sido útil para ilustrar el flujo de un ataque cibernético, pero su aplicación en entornos de IoT ha presentado desafíos debido a la heterogeneidad de los dispositivos y protocolos de IoT, junto con problemas de escala y datos de sensores incompletos o desconectados.

Para abordar estos problemas, se han propuesto y desarrollado nuevos modelos de Kill Chain, adaptados específicamente a los contextos de IoT. Por ejemplo, un nuevo marco de Kill Chain propone una mejor agregación de metadatos y la incorporación de una ontología de amenazas para superar los desafíos presentados por los datos de sensores incompletos o desconectados, como se detalla en el artículo "A Novel Kill-Chain Framework for Remote Security Log Analysis with SIEM Software". Este enfoque mejora la detección de brechas de seguridad y reduce los falsos positivos.

Además, los modelos de Markov, que consideran la naturaleza estocástica de los ciberataques, proporcionan una perspectiva valiosa para comprender y predecir el impacto de los ataques en las organizaciones de IoT, permitiendo también considerar la posibilidad de iteraciones o repetición de ataques.

La seguridad de los dispositivos de IoT es otro tema crucial. La adopción de herramientas de Gestión de Eventos e Información de Seguridad (SIEM), como SecOn, ha demostrado ser eficaz para la protección de sistemas de IoT. Estas herramientas permiten el monitoreo y detección de ataques, ofreciendo una alta precisión y tiempo de respuesta oportuno, como se muestra en el artículo "Testbed-based Evaluation of SIEM Tool for Cyber Kill Chain Model in Power Grid SCADA System".

Otra faceta importante de la cadena de ataque en IoT es la identificación y clasificación de amenazas específicas, como los troyanos de acceso remoto (RAT). En este aspecto, las técnicas de aprendizaje automático y los modelos de caza de amenazas basados en la Cadena de Ataque Cibernética pueden facilitar la detección y prevención de RAT, así como la mitigación de riesgos asociados..

### V. Ataques internos en la computación en la nube

La computación en la nube ha revolucionado la forma en que las empresas almacenan y acceden a la información, ofreciendo escalabilidad, flexibilidad y ahorro de costos significativos. Sin embargo, a medida que más datos se trasladan a la nube, la seguridad de estos recursos se ha convertido en un área crítica de preocupación, y uno de los mayores riesgos en este dominio son los ataques internos.

Los ataques internos en la computación en la nube se refieren a aquellos originados por usuarios legítimos dentro de la misma organización o sistema. Esto puede incluir a empleados descontentos, ex empleados que aún conservan el acceso a los sistemas o incluso a intrusos que han obtenido credenciales de usuario legítimas. Este tipo de atacante puede ser particularmente dañino, ya que ya está dentro del perímetro de seguridad y puede tener un acceso significativo a los recursos del sistema.

Una técnica prometedora para abordar este problema se describe en uno de los resúmenes proporcionados. Este método combina árboles de ataque con la metodología Kill Chain para identificar y detectar ataques internos. Los árboles de ataque son una herramienta que permite visualizar todas las diferentes formas en que un adversario puede lograr un objetivo de ataque. Los nodos representan acciones o eventos, y las aristas representan las relaciones causales entre ellos. Un árbol de ataque completo muestra todas las posibles combinaciones de eventos que conducen a un objetivo de ataque. Cuando se superponen con la Cadena de Ataque, pueden proporcionar una vista detallada de cómo un ataque puede progresar a lo largo de sus diferentes etapas.

Este enfoque puede permitir una detección más efectiva de los ataques internos en la nube al centrarse en los nodos críticos y detectar indirectamente los ataques al identificar precursores y dependencias. Por ejemplo, un ataque interno en la nube podría comenzar con la infiltración de un atacante a través de la explotación de las credenciales de un empleado. El árbol de ataque en este caso podría incluir nodos que representan la obtención inicial de las credenciales, la elevación de privilegios y la exfiltración de datos. La superposición de este árbol de ataque con la Cadena de Ataque podría ayudar a los defensores a identificar las etapas tempranas del ataque, permitiéndoles intervenir antes de que se produzca un daño significativo.

A pesar de las promesas de este enfoque, los desafíos persisten. Los ataques internos pueden ser difíciles de detectar, ya que a menudo se parecen a las actividades normales de los usuarios. Además, la gran cantidad de datos generados en los entornos de la nube puede dificultar la identificación de patrones de ataque sospechosos. Es vital que la investigación futura continúe explorando nuevas técnicas y metodologías para detectar y prevenir estos ataques, salvaguardando así la integridad y seguridad de la computación en la nube.

### VI. Ransomware y la cadena de ataque

El ransomware, una de las amenazas de ciberseguridad más persistentes y perjudiciales en la actualidad, ha cobrado relevancia notablemente en los últimos años. Este tipo de malware opera bajo un mecanismo de cifrado de datos que bloquea a los usuarios el acceso a sus propios archivos o sistemas, y demanda un pago (usualmente en criptomonedas, para mantener el anonimato del atacante) para su liberación. Este tipo de ataque no solo representa una amenaza para la seguridad informática, sino que también plantea importantes desafíos éticos y legales. El crecimiento exponencial en la incidencia de los ataques de ransomware ha transformado este fenómeno en un tema de gran preocupación para gobiernos, empresas y particulares en todo el mundo.

Una herramienta crucial para entender y combatir este tipo de ataques es el modelo de la cadena de ataque o Kill Chain. Este modelo describe las etapas que siguen los atacantes para perpetrar un ciberataque exitoso. Al entender estas etapas, los profesionales de ciberseguridad pueden anticiparse a las tácticas de los atacantes, detectando y neutralizando las amenazas antes de que estas logren su objetivo.
En el caso del ransomware, la cadena de ataque puede ser la siguiente:

1. Reconocimiento: Esta es la etapa inicial en la que los atacantes identifican posibles objetivos y recolectan información que les puede ayudar en etapas posteriores. Por ejemplo, podrían buscar vulnerabilidades en los sistemas de seguridad, entender la estructura de la red, o identificar a empleados que puedan ser susceptibles a tácticas de phishing.
2. Desarrollo y Entrega: Después de recolectar la información necesaria, los atacantes crean el malware y lo entregan al sistema objetivo. Esto puede realizarse mediante una variedad de métodos, que incluyen correos electrónicos de phishing, inyección de código a través de sitios web, o explotación de vulnerabilidades de seguridad en software o hardware.

3. Explotación e Instalación: Una vez que el ransomware ha sido entregado al sistema, explota las vulnerabilidades para instalarse en él.
4. Comando y Control (C2): Después de la instalación, el ransomware se comunica con un servidor controlado por el atacante. Este servidor puede enviar comandos al ransomware y recibir información de él.
5. Acción sobre los Objetivos: Finalmente, el ransomware cifra los datos en el sistema y exige un rescate para descifrarlos.

Este es solo un ejemplo de cómo se puede aplicar la cadena de ataque al ransomware, y las etapas específicas pueden variar según el ataque.

Comprender y aplicar la Kill Chain en el contexto de ataques de ransomware es esencial para desarrollar estrategias de defensa más efectivas. Esta comprensión puede ayudar a identificar puntos débiles en los sistemas de seguridad existentes, desarrollar mejores sistemas de detección de amenazas y formular respuestas más rápidas y efectivas a los ataques.

## VII. BLACKBOARD ARCHITECTURE CYBER COMMAND ENTITY ATTACK PATH IDENTIFICATION (BACCER)

El sistema BACCER opera bajo una arquitectura de pizarra colaborativa, un paradigma innovador que reúne a múltiples expertos en ciberseguridad para trabajar conjuntamente en la detección y mitigación de amenazas cibernéticas. Esta metodología de colaboración en tiempo real facilita un intercambio eficaz de conocimientos y experiencias, permitiendo así la optimización de los recursos disponibles y la creación de un enfoque más sólido y unificado para la ciberseguridad.

BACCER se alimenta de un conjunto diversificado de fuentes de datos, que incluyen registros de eventos de sistemas, registros de red, feeds de inteligencia y bases de datos de vulnerabilidades. Esta recopilación de datos extensiva y heterogénea proporciona una imagen integral y detallada de la actividad en el entorno objetivo, facilitando la detección temprana de actividades sospechosas o amenazas emergentes.

Una vez recopilados los datos, BACCER los analiza utilizando un conjunto de reglas predefinidas que describen comportamientos maliciosos conocidos. Cuando las condiciones de una regla se cumplen, se genera una alerta inmediata, proporcionando una indicación temprana de una posible actividad maliciosa en progreso.

No obstante, BACCER va más allá de un simple sistema de correspondencia de reglas. Recopila y procesa hechos en tiempo real, proporcionando un contexto más detallado y actualizado de la situación de seguridad cibernética actual. Esto puede incluir elementos como direcciones IP sospechosas, comportamientos inusuales de usuarios, o intentos de acceso no autorizado, que se utilizan en combinación con las reglas para mejorar la toma de decisiones.

La capacidad de BACCER para evaluar amenazas se basa en una lógica de decisión robusta que evalúa la gravedad y el impacto potencial de cada amenaza identificada. A través de algoritmos avanzados, este sistema asigna un nivel de riesgo a cada patrón de ataque identificado, permitiendo así una priorización y respuesta más precisa y efectiva a las amenazas.

Las contramedidas activadas por BACCER son tan variadas como efectivas e incluyen el bloqueo de direcciones IP sospechosas, el aislamiento de sistemas comprometidos y la generación de informes detallados para la revisión y el análisis por parte de expertos humanos. Estas respuestas se basan en las mejores prácticas y políticas de seguridad actuales, garantizando una mitigación de amenazas eficaz y en línea con los estándares de la industria.

En resumen, el sistema BACCER representa un enfoque proactivo y holístico para la ciberseguridad. Mediante la combinación de una arquitectura de pizarra colaborativa, el análisis de datos basado en reglas y hechos, una lógica de decisión bien afinada y la activación de contramedidas eficaces, BACCER está posicionado para fortalecer las defensas cibernéticas, prevenir ataques y minimizar los daños, protegiendo así nuestros activos digitales en un mundo cada vez más conectado.

La evaluación de la eficacia del sistema BACCER se basa en su capacidad para identificar y mitigar los ataques cibernéticos mediante el análisis de árboles de ataque y defensa, así como cadenas de ataque a la red. A continuación, te proporciono una explicación sobre cómo se evalúa y algunos ejemplos de su eficacia:

**Evaluación de la eficacia:**

- **Precisión de detección:** Se evalúa la capacidad del sistema BACCER para detectar y alertar sobre patrones de ataque reales sin generar falsos positivos. Esto implica comparar las alertas generadas por el sistema con los ataques confirmados o eventos maliciosos reales.
- **Tiempo de respuesta:** Se mide la velocidad con la que el sistema BACCER responde a las amenazas detectadas. Un tiempo de respuesta rápido es esencial para mitigar los ataques antes de que causen daños significativos.
- **Cobertura de amenazas:** Se evalúa la capacidad del sistema para abordar una amplia gama de amenazas cibernéticas, incluyendo ataques conocidos y nuevos patrones de ataque emergentes.
- **Capacidad de adaptación:** Se analiza la capacidad de BACCER para adaptarse y actualizar su conocimiento y reglas en tiempo real a medida que evolucionan las amenazas y surgen nuevas tácticas y técnicas de ataque.

**Ejemplos de eficacia:**

- Identificación de ataques avanzados: BACCER es capaz de detectar y alertar sobre ataques avanzados que involucran múltiples etapas y técnicas sofisticadas, como ataques de inyección SQL, ataques de phishing dirigidos o ataques de ransomware.
- **Prevención de brechas de seguridad:** Mediante el análisis de árboles de ataque y defensa, BACCER puede identificar posibles brechas de seguridad y proporcionar contramedidas efectivas para prevenir la explotación exitosa de esas brechas.
- **Interrupción de cadenas de ataque:** Al utilizar la Cadena de Ataque (Cyber Kill Chain), BACCER puede identificar y romper las diferentes etapas de un ataque cibernético, como el reconocimiento, la explotación y la persistencia, impidiendo así el avance del atacante en la red.
- **Colaboración efectiva:** BACCER fomenta la colaboración entre expertos en ciberseguridad al proporcionar una plataforma compartida para el análisis de amenazas. Esta colaboración mejora la eficacia del sistema al aprovechar la experiencia colectiva y la retroalimentación de los expertos.

En general, la eficacia del sistema BACCER se evalúa en términos de su capacidad para identificar y responder de manera efectiva a los patrones de ataque y defender las redes contra amenazas cibernéticas. Su enfoque colaborativo, combinado con el análisis de árboles de ataque y defensa y cadenas de ataque, lo convierte en una herramienta valiosa en la lucha contra los ataques cibernéticos.

## VIII. Implicaciones y futuras direcciones

El auge del ransomware y otras amenazas avanzadas de ciberseguridad tiene implicaciones profundas y de largo alcance. Con las tecnologías modernas cada vez más integradas en nuestras vidas cotidianas, tanto a nivel individual como empresarial, la seguridad de nuestros sistemas digitales ha adquirido una importancia primordial. Los ataques de ransomware a sistemas ciber físicos, a la banca online mediante troyanos y a los sistemas de control industrial representan amenazas significativas no sólo para la integridad de los datos, sino también para la infraestructura física, la economía y la seguridad nacional.

A medida que estas amenazas evolucionan, las estrategias de ciberseguridad deben mantenerse al día. La inteligencia computacional evolutiva, por ejemplo, puede utilizarse para adaptar y mejorar los sistemas de defensa en tiempo real, permitiendo a las defensas "aprender" de los ataques pasados y adaptarse a las nuevas amenazas a medida que surgen. Esta técnica puede proporcionar una capa adicional de protección al mejorar la detección y la respuesta a los ataques.

La identificación de escenarios de amenazas mediante gráficos de ataque también puede mejorar la defensa contra las amenazas. Estos gráficos, que representan las posibles rutas que un atacante puede seguir a través de una red, pueden ayudar a los defensores a anticipar y prepararse para los movimientos del atacante. Esto puede mejorar tanto la prevención de los ataques como la velocidad y la eficacia de la respuesta cuando se producen.

El análisis de la cadena de ataque también tiene un papel crucial en la mejora de la seguridad, y su aplicación puede ser ampliada más allá de los contextos en los que se ha utilizado tradicionalmente. Por ejemplo, la cadena de ataque puede aplicarse al análisis de la seguridad de los servicios multimedia, proporcionando un marco para evaluar y mejorar la seguridad en este sector. La cadena de ataque también puede utilizarse para estudiar y mejorar la defensa contra los troyanos bancarios y otras amenazas a los sistemas financieros.

Las tácticas, técnicas y procedimientos utilizados por los actores de amenazas deben estudiarse y entenderse para mejorar las defensas. La información obtenida de estos estudios puede utilizarse para proponer nuevos métodos y modelos de ciberdefensa, mejorar la detección de amenazas y reforzar la seguridad de nuestros sistemas.

En resumen, las implicaciones de las amenazas avanzadas de ciberseguridad son profundas y variadas, y requieren un enfoque amplio y adaptable para la defensa. La evolución continua de las tácticas y las técnicas de ataque requiere una respuesta igualmente evolutiva por parte de los defensores, y técnicas como la inteligencia computacional evolutiva, los gráficos de ataque y la aplicación de la cadena de ataque a nuevos contextos pueden ser partes fundamentales de esa respuesta. Al explorar y adoptar estas y otras técnicas avanzadas de ciberseguridad, podemos mejorar la seguridad y la resiliencia de nuestros sistemas digitales ahora y en el futuro.

## IX. Conclusiones

- Los modelos de Kill Chain, como el Modelo de Kill Chain de Lockheed Martin y el Modelo MITRE ATT&CK, ofrecen una estructura sólida para analizar y defenderse contra los ataques cibernéticos, permitiendo comprender las etapas seguidas por los atacantes y desarrollar estrategias adaptativas de mitigación y respuesta.
- La seguridad de la computación en la nube se ha vuelto crítica debido al traslado masivo de datos a este entorno, y abordar los ataques internos es fundamental para proteger los recursos y la información de las organizaciones.
- La comprensión y aplicación de la cadena de ataque o Kill Chain en el contexto del ransomware es fundamental para desarrollar estrategias de defensa efectivas. Al entender las etapas que siguen los atacantes, los profesionales de ciberseguridad pueden anticiparse a sus tácticas, detectar y neutralizar las amenazas antes de que logren su objetivo.
- La ciberseguridad se ha convertido en un desafío crucial en el entorno digital actual, y es necesario contar con herramientas y enfoques eficaces para proteger nuestros sistemas, datos y activos. Tanto el conocimiento de la cadena de ataque como la implementación de sistemas como BACCER representan pasos importantes en la dirección correcta para garantizar la seguridad y la integridad de nuestras infraestructuras digitales.
- La inteligencia computacional evolutiva y los gráficos de ataque son técnicas emergentes que pueden mejorar significativamente la detección y la respuesta a las amenazas. Al adaptarse a las nuevas amenazas y entender mejor cómo se mueven los atacantes a través de una red, estas técnicas pueden mejorar la eficacia de nuestras defensas.

## REFERENCIAS

[1] P. N. Bahrami, A. Dehghantanha, T. Dargahi, R. M. Parizi, K. K. R. Choo, and H. H. S. Javadi, "Cyber Kill Chain-Based Taxonomy of Advanced Persistent Threat Actors: Analogy of Tactics, Techniques, and Procedures," Journal of Information Processing Systems, vol. 15, no. 4, pp. 865-889, 2019.

[2] D. Kiwia, A. Dehghantanha, K.-K. R. Choo, and J. Slaughter, "A Cyber Kill Chain Based Taxonomy of Banking Trojans for Evolutionary Computational Intelligence," Journal of Computational Science, vol. 27, pp. 394-409, 2018.

[3] A. Hahn, R. K. Thomas, I. Lozano, and A. Cardenas, "A Multi-layered and Kill-chain Based Security Analysis Framework for Cyber-Physical Systems," International Journal of Critical Infrastructure Protection, vol. 11, pp. 39-50, 2015.

[4] P. Liu, L. Zhu, Z. Zhou, and H. Xiao, "Research on Kill Chain Analysis Method Based on Template-Bayesian Network," in 2021 2nd International Conference on Electronics, Communications and Information Technology (CECIT), 2021, p. 5.

[5] L. Sadlek, P. Celeda, and D. Tovarnak, "Identification of Attack Paths Using Kill Chain and Attack Graphs," in NOMS 2022-2022 IEEE/IFIP Network Operations and Management Symposium, 2022.

[6] Y. Wang, C. Wang, S. Liu, and X. Yan, "Kill chain for industrial control system," in MATEC Web of Conferences, vol. 246, 2018, Art. no. 01013.

[7] A. Kott, C. Wang, and R. F. Erbacher, Eds., Cyber Defense and Situational Awareness. Advances in Information Security. Springer International Publishing, 2014.

[8] S. Cho, I. Han, H. Jeong, J. Kim, S. Koo, H. Oh, and M. Park, "Cyber Kill Chain based Threat Taxonomy and its Application on Cyber Common Operational Picture," IEEE Xplore, Jun. 1, 2018.

[9] S. K. Ghosh and A. Kumar, "Modified Cyber Kill Chain Model for Multimedia Service Environments," Multimedia Tools and Applications, vol. 77, no. 20, pp. 27207-27229, 2018.

[10] J. Haseeb, M. Mansoori, and I. Welch, "A Measurement Study of IoT-Based Attacks Using IoT Kill Chain," IEEE.

[11] J. Straub, "Modeling Attack, Defense and Threat Trees and the Cyber Kill Chain, ATT&CK and STRIDE Frameworks as Blackboard Architecture Networks," IEEE.

[12] N. Naik, J. Song, P. Grace, and P. Jenkins, "Comparing Attack Models for IT Systems: Lockheed Martin's Cyber Kill Chain, MITRE ATT&CK Framework and Diamond Model," IEEE.

[13] A. Duncan, M. Goldsmith, and S. Creese, "A Combined Attack-Tree and Kill-Chain Approach to Designing Attack-Detection Strategies for Malicious Insiders in Cloud Computing," IEEE.

[14] A. Duncan, M. Goldsmith, y S. Creese, "A Combined Attack-Tree and Kill-Chain Approach to Designing Attack-Detection Strategies for Malicious Insiders in Cloud Computing," en IEEE, s.f., [En línea]. Disponible en: https://ieeexplore-ieee org.bdigital.udistrital.edu.co/document/8885401

[15] Q. K. Ali Mirza, A. Alam, L. Shand, O. Halling, y M. Brown, "Ransomware Analysis using Cyber Kill Chain", en IEEE, [En línea]. Disponible en: https://ieeexplore-ieee-org.bdigital.udistrital.edu.co/document/9590453

[16] T. Neubert and C. Vielhauer, "Kill chain attack modelling for hidden channel attack scenarios in industrial control systems," in IFAC-PapersOnLine, vol. 53, no. 2, pp. 11074-11080, 2020. Available: https://doi.org/10.1016/j.ifacol.2020.12.246.

[17] A. T. Ahmed, M. Arafatur Rahman, "A Cyber Kill Chain Approach for Detecting Advanced Persistent Threats," Computers, Materials & Continua, vol. 67, no. 2, pp. 2497-2513, 2021. [Online]. Available: https://doi.org/10.32604/cmc.2021.014223.

[18] P. Zanna, Radcliffe, and D. Kumar, "Preventing Attacks on Wireless Networks Using SDN Controlled OODA Loops and Cyber Kill Chains," Sensors (Basel, Switzerland), vol. 22, no. 23, pp. 9481–, 2022. [Online]. Available: https://doi.org/10.3390/s22239481

[19] A. Dargahi, A. Dehghantanha, P. N. Bahrami, M. Conti, G. Bianchi, and L. Benedetto, "A Cyber-Kill-Chain based taxonomy of crypto-ransomware features," Journal of Computer Virology and Hacking Techniques, vol. 15, no. 4, pp. 277–305, 2019. [Online]. Available: https://doi.org/10.1007/s11416-019-00338-7

[20] Ullah, Y., Ullah, M., Ullah, H., Katt, B., Hijji, M., & Muhammad, K. (2022). Mapping Tools for Open Source Intelligence with Cyber Kill Chain for Adversarial Aware Security. Mathematics (Basel), 10(12), 2054-.

[21] B. D. Bryant and H. Saiedian, "A novel kill-chain framework for remote security log analysis with SIEM software," Computers & Security, vol. 67, pp. 198-210, 2017. [Online]. Available: https://doi.org/10.1016/j.cose.2017.03.003.

[22] R. Hoffmann, "Markov Models of Cyber Kill Chains with Iterations," in International Conference on Military Communications and Information Systems, 2019, pp. 6, doi: 10.1109/ICMCIS.2019.8842810.

[23] J. M. Spring and E. Hatleback, "Thinking about intrusion kill chains as mechanisms," J. Cybersecur., vol. 3, no. 1, p. tyw012, 2017.

[24] V. Kumar, S. Perez, and M. Govindarasu, "Testbed-based Evaluation of SIEM Tool for Cyber Kill Chain Model in Power Grid SCADA System," in North American Power Symposium, 2019, pp. 6, https://doi.org/10.1109/NAPS46351.2019.9000344.

[25] R. HosseiniNejad, H. HaddadPajouh, A. Dehghantanha, R.M. Parizi (2019). "A Cyber Kill Chain Based Analysis of Remote Access Trojans." In: A. Dehghantanha, KK. Choo (eds), Handbook of Big Data and IoT Security, pp. 12. Springer, Cham.

[26] M. A. Khan and S. U. Khan, "A model-driven approach for early analysis of kill chain and resource starvation attack in MQTT-based IoT systems," Journal of Ambient Intelligence and Humanized Computing, vol. 11, no. 7, pp. 2925-2936, 2020. [Online]. Available: https://doi.org/10.1007/s12652-020-01747-1

[27] A. Villalón-Huerta, P. García-Teodoro, G. Maciá-Fernández, and K. Gil-González, "SOC Critical Path: A Defensive Kill Chain Model," in IEEE Access, vol. 10, pp. 13581-13594, 2022. doi: 10.1109/ACCESS.2022.3177645.

[28] F. E. Salamh, U. Karabiyik, M. K. Rogers, and E. T. Matson, "Unmanned Aerial Vehicle Kill Chain: Purple Teaming Tactics," in IEEE Access, vol. 7, pp. 107982-107994, 2019, doi: 10.1109/ACCESS.2019.2939262.

[29] A. Duncan, S. Creese, y M. Goldsmith, "A combined attack-tree and kill-chain approach to designing attack-detection strategies for malicious insiders in cloud computing," en 2019 International Conference on Cyber Security and Protection of Digital Services (Cyber Security).

[30] M. S. Kumar, G. U. Krishnan, J. Ben-Othman, and K. G. Srinivasagan, "Artificial Intelligence Managed Network Defense System against Port Scanning Outbreaks," in 2019 International Conference on Vision Towards Emerging Trends in Communication and Networking (ViTECoN), pp. 1-5, IEEE.